\documentclass[twocolumn]{aastex62}

\received{}
\revised{}
\accepted{}

\shorttitle{DES-unWISE $z \gtrsim 6.5$ Quasar Survey}
\shortauthors{Yang et al.}

\begin{document}

\title{Exploring Reionization-Era Quasars IV: Discovery of Six New $z \gtrsim 6.5$ Quasars with DES, VHS and unWISE Photometry}

\correspondingauthor{Jinyi Yang}
\email{jinyiyang@email.arizona.edu}

\author{Jinyi Yang}
\affil{Steward Observatory, University of Arizona, 933 N Cherry Ave, Tucson, AZ, USA}

\author{Feige Wang}
\affil{Department of Physics, University of California, Santa Barbara, CA 93106-9530, USA}

\author{Xiaohui Fan}
\affil{Steward Observatory, University of Arizona, 933 N Cherry Ave, Tucson, AZ, USA}

\author{Minghao Yue}
\affil{Steward Observatory, University of Arizona, 933 N Cherry Ave, Tucson, AZ, USA}

\author{Xue-Bing Wu}
\affil{Kavli Institute for Astronomy and Astrophysics, Peking University, Beijing 100871, China}
\affil{Department of Astronomy, School of Physics, Peking University, Beijing 100871, China}

\author{Jiang-Tao Li}
\affil{Department of Astronomy, University of Michigan, 311 West Hall, 1085 S. University Ave, Ann Arbor, MI, 48109-1107, USA}

\author{Fuyan Bian}
\affil{European Southern Observatory, Alonso de C\'ordova 3107, Casilla 19001, Vitacura, Santiago 19, Chile}

\author{Linhua Jiang}
\affil{Kavli Institute for Astronomy and Astrophysics, Peking University, Beijing 100871, China}

\author{Eduardo Ba{\~n}ados}
\affil{The Observatories of the Carnegie Institution for Science, 813 Santa Barbara Street, Pasadena, California 91101,USA}

\author{Yuri Beletsky}
\affil{The Observatories of the Carnegie Institution for Science, 813 Santa Barbara Street, Pasadena, California 91101,USA}



\begin{abstract}
This is the fourth paper in a series of publications aiming at discovering quasars at the epoch of reionization. In this paper, we expand our search for $z\sim 7$ quasars to the footprint of the Dark Energy Survey (DES) Data Release One (DR1), covering $\sim 5000$ deg$^2$ of new area. We select $z\sim 7$ quasar candidates using deep optical, near-infrared(near-IR) and mid-IR photometric data from the DES DR1, the VISTA Hemisphere Survey (VHS), the VISTA Kilo-degree Infrared Galaxy (VIKING) survey, the UKIRT InfraRed Deep Sky Surveys -- Large Area Survey (ULAS) and the unblurred coadds from the {\it Wide-field Infrared Survey Explore} ({\it WISE}) images (unWISE). The inclusion of DES and unWISE photometry allows the search to reach $\sim$ 1 magnitude fainter, comparing to our $z \gtrsim 6.5$ quasar survey in the northern sky \citep{wang18b}. We report the initial discovery and spectroscopic confirmation of six new luminous quasars at $z>6.4$, including an object at $z=7.02$, the fourth quasar yet known at $z>7$, from a small fraction of candidates observed thus far. Based on the recent measurement of $z \sim 6.7 $ quasar luminosity function using the quasar sample from our survey in the northern sky, we estimate that there will be $\gtrsim$ 55 quasars at $z > 6.5$ at $M_{1450} < -24.5$ in the full DES footprint.

\end{abstract}



\section{Introduction} \label{sec:intro}
Luminous quasars at high redshift provide direct probes of the evolution of supermassive black holes (BHs) and the intergalactic medium (IGM) at early cosmic time. Over the last decade, more than 150 quasars have been discovered at $z > 6$ \citep[e.g.][]{fan01,fan06,willott10b,mortlock11,venemans13,venemans15,kashikawa15,wu15,matsuoka16,jiang16,wang17,mazzucchelli17,reed15, reed17, banados16,banados18,matsuoka18a,matsuoka18b}, with the highest redshift at $z = 7.54$ \citep{banados18}. Detections of such objects indicate the existence of billion $M_{\odot}$ BHs merely a few hundred Myrs after the first star formation in the Universe. It challenges the theory of BH growth and places the strongest constraints on the BH-galaxy coevolution at early epoch \citep[e.g.][]{wu15}.
Absorption spectra of the highest redshift quasars reveal complete Gunn-Peterson absorption, indicating a rapid increase in the IGM neutral fraction, marking the end of the reionization epoch at $z > 6$ \citep[see][]{fan06}. Combined with results from the most recent detections on the declining Ly$\alpha$ visibility and abundance among high-redshift galaxies and revised measurements of Thompson optical depth from CMB polarization, current data strongly suggest a peak of reionization activity and emergence of the earliest galaxies and AGNs at $6 < z < 10$ \citep{robertson15, planck18}, highlighting the need to expand our search to higher redshift.

At $z \gtrsim 7$, Ly$\alpha$ emission line is redshifted to beyond 1 micron, requiring near-IR observations for both their selections and spectroscopic identifications. Efforts to find such objects have proven to be difficult. Until recently, only two quasars have been discovered at $z \gtrsim 7$, and a dozen at $z > 6.5$, compared to the large number of available objects at $z \sim 6$, despite decade long effort by the high-redshift quasar community. This is due to a combination of their declining spatial density, limited sky coverage of near-IR photometry, and low efficiency in spectroscopic follow-up observations.
This situation is rapidly changing, with new deep optical and infrared sky survey data finally becoming available. 
In the optical, the Dark Energy Spectroscopic Instrument (DESI) Legacy Imaging Surveys \citep[DELS,][]{dey18} is covering the SDSS footprint with high quality, deep photometry in $g, r$ and $z$ bands, the Pan-STARRS1 \citep[PS1,][]{chambers16} survey provides $g, r, i, z$ and $y$ observations over 3-$\pi$ area of the sky and the Dark Energy Survey \citep[DES,][]{abbott18} DR1 reaches even fainter flux level in $g, r, i, z$ and $Y$ bands over 5000 deg$^{2}$ in the southern sky; in the near-IR, the UKIRT Hemisphere Survey (UHS) \citep{dye18}, the UKIRT InfraRed Deep Sky Surveys--Large Area Survey \citep[ULAS,][]{lawrence07}, the VISTA Hemisphere Survey \citep[VHS,][]{mcmahon13} and the VISTA Kilo-degree Infrared Galaxy survey \citep[VIKING,][]{edge13} are mapping almost the entire sky; in the mid-IR, the latest {\it Wide-field Infrared Survey Explore} \citep[{\it WISE},][]{wright10} survey from the Near-Earth Object Wide-field Infrared Survey Explorer \citep[NEOWISE,][]{mainzer11,mainzer14} project has doubled the depth in the critical W1 and W2 bands. The unblurred coadds of {\it WISE} \citep[unWISE,][]{lang14} have better resolution and enhanced depths than blurred ALLWISE data in the W1/W2 bands with new releases of the  additional data from the NEOWISE-Reactivation \citep[NEOWISER][]{mainzer14} phase of the mission \citep{meisner17,meisner18}.

The combination of these surveys allows the optical/near-IR selection of quasars at $z \gtrsim 6.5$ in deeper and wider field sky than in the past. Our group is conducting a survey of quasars at $z \gtrsim 6.5$ \citep{fan18,wang17,wang18a,wang18b} utilizing these new datasets. \cite[][hereafter Paper III]{wang18b} presents the discovery of 16 new quasars from our survey in the northern sky using the combination of DELS, PS1, UHS, ULAS, VHS, VIKING and {\it WISE} data; we have doubled the number of known quasar at $z > 6.5$, measured the quasar luminosity function (QLF) at $z = 6.7$ and constrained the quasar number density evolution during the reionzation epoch. In this paper, we present our on-going survey in the southern sky based on similar optical/IR selection method, using the data from DES, VHS, VIKING and unWISE, and report the initial discoveries of six new quasars at $6.41 \le z \le 7.02$ based on a small fraction of candidates observed. We introduce the photometric dataset and quasar selection criteria used in this survey in Section 2. The spectroscopic observations and the result are described in Section 3 and Section 4. We discuss the implication of these discoveries and conclude with a short summary in Section 5. In this paper, we adopt a $\Lambda$CDM cosmology with parameters $\Omega_{\Lambda}$ = 0.7, $\Omega_{m}$ = 0.3, and H$_{0}$ = 70 $km s^{-1} Mpc^{-1}$. 
Photometric data from DES are reported on the AB system after applying the Galactic extinction correction \citep{schlegel98,schlafly11}; photometric data from IR (e.g. VHS, VIKING, ULAS and WISE) surveys are reported in the Vega system. 

\section{Candidate Selection} \label{sec:sel}
\subsection{Photometric Datasets}

\begin{table}
\caption{Photometric data used in this survey.}
\scriptsize
\begin{tabular}{l l l l}
\hline\hline
  \multicolumn{1}{c}{Survey} &
  \multicolumn{1}{c}{Area ($deg^{2}$)} &
  \multicolumn{1}{c}{Data used} &
  \multicolumn{1}{c}{Depth (5$\sigma$)} \\
\hline
  DES DR1& 5000 & g, r, i, z, Y & 25.08, 24.83, 24.19, 23.44, 22.19\\
  VHS DR5\tablenotemark{a} & 3600 & J, Ks & 20.2, 19.2, 18.2 (Vega) \\
  VIKING DR4 & 300 & J, Ks & 21.2, 19.4 (Vega) \\
  ULAS DR10 & 170 & J, K & 19.6, 18.2 (Vega) \\
  unWISE & all-sky & W1, W2 & 18.3\tablenotemark{b}, 16.9 (Vega)\\
\hline
\end{tabular}
\tablenotetext{a}{\scriptsize We only show the near-IR -- DES overlapped area here.}
\tablenotetext{b}{\scriptsize The unWISE depth is estimated from the magnitude-error relation of unWISE data in the DES area.}
\end{table}

This survey is a part of our wide field $z \gtrsim 6.5$ quasar survey based on optical/IR photometry. Similar to our survey in the northern sky in Paper III, we are mainly using colors in $z, Y, J$ and W1 bands but modify the selection criteria according to the DES filters. The photometric data is from DES DR1, VHS, VIKING, ULAS and unWISE. Except for the $\sim 170$ deg$^{2}$ field that is only covered by ULAS without VHS images, photometric data used here is deeper than the optical/IR data used for our survey in the northern sky, which enables us to search quasars with $\gtrsim$ 1 magnitude fainter.
A summary of photometric data used in this survey is shown in Table 1.

In the optical, we use DES DR1 photometric data.  
The DES survey uses the Dark Energy Camera \citep[DECam,][]{honscheid08, flaugher15} on the CTIO 4-m Blanco Telescope to image five broad bands ($g, r, i, z$ and $Y$) in a 5000 deg$^{2}$ field in the southern Galactic cap. The DR1 data encompasses images from first three years of the survey, with a coadd magnitude limit (MAG\_APER\_4, 1.95 arcsec diameter,10 $\sigma$) of 24.33, 24.08, 23.44, 22.69 and 21.44 in the five bands, respectively. 
In particular, the DES $Y$ band covers the wavelength from 9300 \AA\ to 10700 \AA\, extended to the near-IR wavelength range.
It reaches $\sim$ 1 magnitude fainter than $z $ PS1 survey, 0.6 magnitude fainter than $z$ in DELS (DR4+DR5), and covers redder wavelength range ($\sim$ 700 \AA\ ) than PS1 $z$. DES $z$ imaging allows the detection of quasar at redshift up to $z \sim 7.2$.

In the near-IR, we are using all public near-IR images in the DES area, including those from the VHS, VIKING and ULAS. The VHS survey aims at imaging the entire hemisphere of the southern sky in the $Y, J, H$ and $Ks$ bands, with the depth (5 $\sigma$) to 20.6, 20.2, 19.2 and 18.2 mag in Vega. The VHS--DES area will be $\sim 4400$ deg$^{2}$, while the current VHS DR5 has covered $\sim 80$\% of the entire VHS--DES area. The VIKING survey covers $\sim$ 1500 deg$^{2}$ in $Z, Y, J, H$ and $Ks$ bands over three extragalactic areas, with the 5 $\sigma$ depth of 22.6, 21.7, 21.2, 20.1, 19.4 mag. The entire VIKING--DES area is about 500 deg$^{2}$, of which about 300 deg$^{2}$ is available now. The ULAS survey has $\sim 450$ deg$^{2}$ overlap region with the DES, among which $\sim$ 280 deg$^{2}$ field is also covered by the VHS. For the area covered by both VHS and ULAS we use the VHS data as it is deeper than ULAS. The survey depth of ULAS in $Y, J, H$ and $K$ bands is 20.2, 19.6, 18.8 and 18.2 mag. 

Discoveries presented in this paper used DES $Y$ band as the detection band.
To improve the signal-to-noise (S/N) ratio of faint objects, we carried out forced photometry of each object in $J$ and $Ks$ ($K$) band centered on the position from DES DR1. We used a 2 arcsec aperture diameter, the same to that used for $J$AperMag3 in VHS, VIKING and ULAS surveys. The forced photometry also provides measurements for faint objects that are covered by near-IR imaging but are below the detection limit of the public catalog.  The $J$ and $Ks$($K$) band magnitudes used in the color selection in Section 2.2 are our forced photometric magnitudes for all objects.

Photometric data in {\it WISE} W1 and W2 bands is useful to separate quasar from brown dwarfs (e.g. the W1-W2 color). Compared with the deep DES, VHS and VIKING survey, the ALLWISE data with the depth (5 $\sigma$) of 17.6 and 16.2 mag is relative shallow, with position-dependent depth (coverage). Therefore, we performed forced photometry on the unWISE coadded image of each object using the DES position. We used the forced photometry code from {\it The Tractor} \citep{lang16a, lang16b}. The current unWISE coadded image incorporates {\it WISE} and three years of NEOWISER imaging. The forced photometric unWISE data in the DES area is $\sim$ 0.7 magnitude deeper than ALLWISE, based on the magnitude-error relation of the forced photometric data in the DES area. 

\subsection{Selection Criteria}
\begin{figure}
\centering
\epsscale{1.0}
\plotone{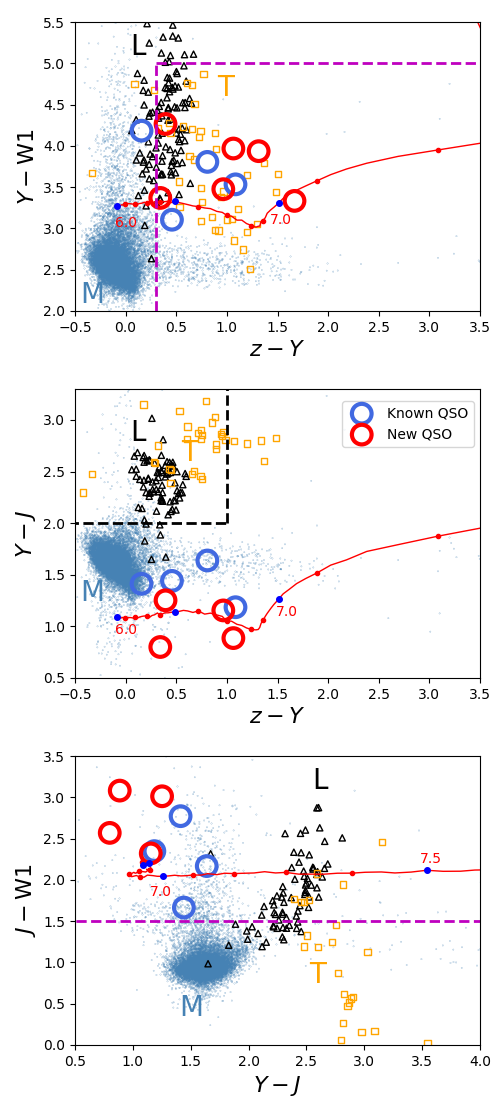} 
\caption{The quasar selection color-color cuts, in $z-Y$/$Y-$ W1, ({\it Top}) $z-Y/Y-J$ ({\it Middle}) and $Y-J/J-$ W1 ({\it Bottom}) color-color diagrams. The red solid lines with red/blue points are the quasar color track started at $z = 6.0$ with the step of $\Delta z = 0.1$. The blue points mark the redshifts of $z = 6.0, 6.5, 7.0$ and 7.5. The blue open circles represent colors of four previously known $z \ge 6.5$ quasars in the DES DR1. The red open circles denote six new quasars presented in this paper. For $J$ and $W1$, forced photometry measurements are used. Quasars J025216--050331 and J031941--100846 do not have near-IR images and thus there are only four red open circles in the $z-Y/Y-J$ and $Y-J/J-$ W1 diagrams. The purple/black dashed lines represent the selection/rejection criteria.   
} 
\end{figure}

We started our selection with the DES DR1 catalog \citep{abbott18} and used the MAG\_APER\_4 magnitude. We first selected objects that had $> 5 \sigma$ detection in $Y$ band but were undetected in $g$ and $r$ band with the S/N $<$ 3. We require the objects to be undetected in $i$ band (S/N($i$) $<$ 3) or with $i-z > 3$ color. The former cut is used to select the fully $i$ band drop-out objects (we call it $'$main sample$'$), and the latter one is designed for ultra-luminous quasars which are still detected in the deep DES $i$ band ($'$luminous sample$'$).  
We used the baseline quality criterion of IMAFLAGS\_ISO = 0 \citep{abbott18} for DES photometry. 
We selected objects in the $z-Y/Y-$W1 color space for quasars at $z \gtrsim 6.5 $ and reject most of M dwarfs and part of L, T dwarfs, as shown in Figure 1.
Then we cross-matched (2$''$ radius) color selected candidates with PS1 DR1 and did forced photometry in PS1 $g ,r, i$ and $z$ bands using 2$''$ diameter aperture to further reduce candidates. We reject objects with $i_{ps1,forced}$ brighter than 23.1 mag or $z_{ps1,forced}$ brighter than 22.3 mag in the main sample, and only limit the $i_{ps1,forced}$ for objects in the luminous sample. The selection criteria are listed as following.
\begin{equation}
S/N(g)< 3.0, S/N(r)< 3.0, S/N(Y) > 5.0
\end{equation}
\begin{equation}
 \begin{array}{l}
Main: S/N(i) < 3.0 \\
Luminous: S/N(i) >= 3.0, i-z > 3.0
\end{array}
\end{equation}
\begin{equation}
z - Y > 0.3 ~ and ~ Y - W1 < 6
\end{equation}
\begin{equation}
0 < W1 - W2 < 2
\end{equation}
\begin{equation}
 \begin{array}{l}
Main: i_{ps1,forced} > 23.1 ~and~ z_{ps1,forced} > 22.3 \\
Luminous: i_{ps1,forced} > 23.1
 \end{array}
\end{equation}

To further reject contaminants, we add near-IR $J$ and $Ks$ ($K$) band data from the VHS, VIKING and ULAS based on forced photometry measurements described above.  
Since the near-IR surveys do not fully cover the DES area, we kept all objects that do not have $J$ or $Ks$ images. We used the $z-Y/Y-J$ and $J-$ W1 colors for objects rejection, as shown in Figure 1. We rejected all objects which had $J$ band image but with negative flux from the forced photometry.
A $J - Ks/K$ cut was added for objects that had $Ks$($K$) detection, which is designed for further rejection based on the power law continuum of quasar. We {\it rejected} objects that satisfied any one criterion listed below.

\begin{equation}
z-Y < 1~and~ Y-J> 2
\end{equation}
\begin{equation}
J - W1 < 1.5
\end{equation}
\begin{equation}
J - Ks < 0.5
\end{equation}

After applying all selection criteria above, we visually inspected optical/IR image of each candidate. We rejected extended objects, objects visible in any of DES $g, r, i$ (except those with $i$-band detection in the 'luminous sample') and PS1 $g, r, i$ bands, and objects with photometry contaminated by cosmic rays or nearby bright stars. We finally obtained a candidate sample including $\sim 380$ candidates, 350 from main sample and 30 from luminous sample. In the DES area, there are five previously known quasars at $z \ge 6.5$ \citep{venemans13,venemans15,reed17,matsuoka18a}. Four quasars are included in the DES DR1 catalog, except for the faint quasar HSCJ0213--0626 \citep{matsuoka18a}. Among these four known quasars, three are covered by our selection: J010953--304736 at $z =6.79$, J030516--315056 at $z=6.61$ \citep{venemans13} and PSO J036+03 (also named as J022601.873+030259.254) \citep{venemans15}. The quasar J022426--471129 at $z=6.5$ from \cite{reed17} was rejected by the $z-Y>0.3$ cut. 

\begin{figure}
\centering
\epsscale{1.2}
\plotone{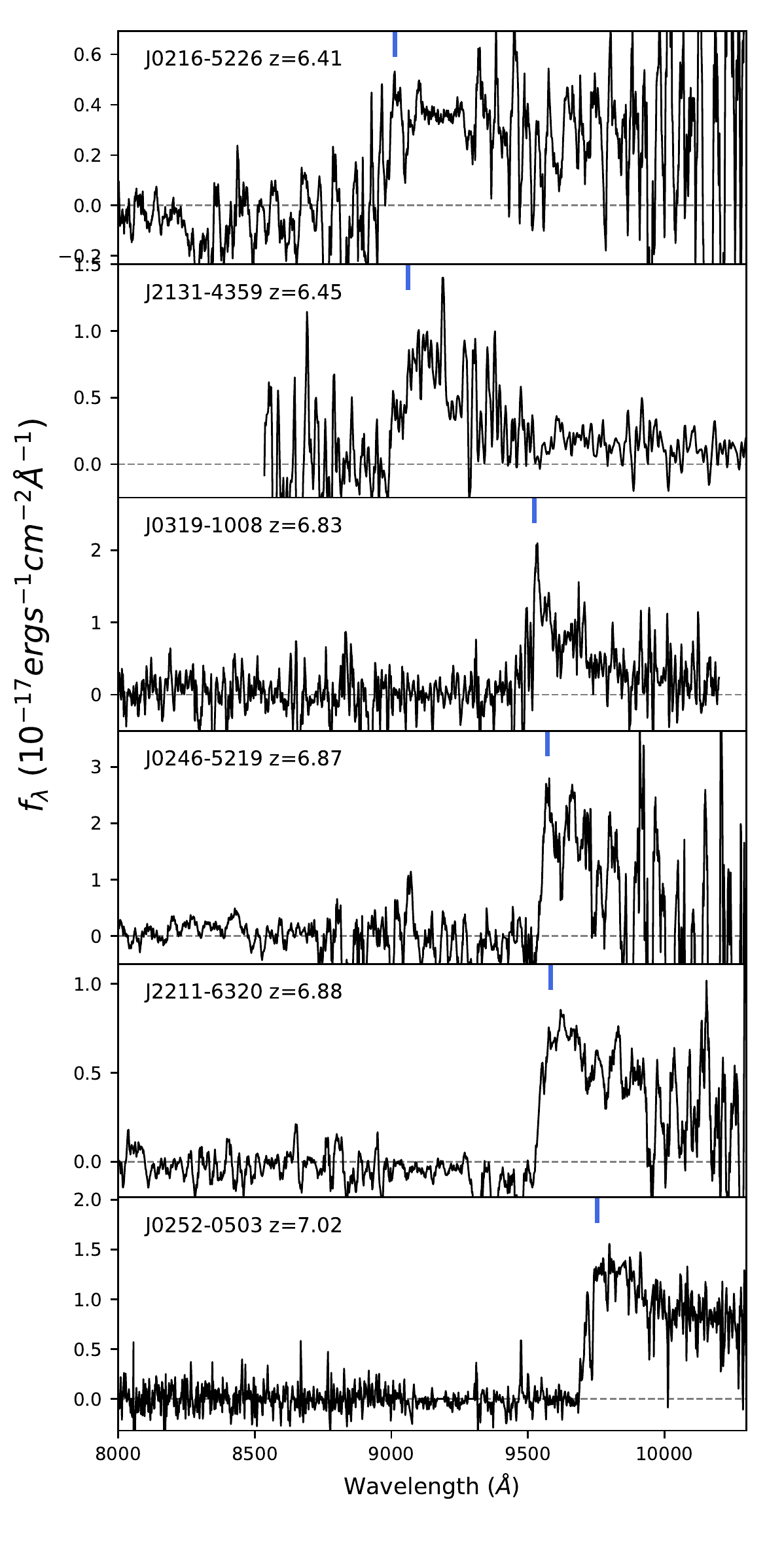} 
\caption{Spectra of the six new $z>6.4$ quasars. The blue vertical lines mark the locations the $\rm Ly \alpha$ emission line. Spectra from Magellan/LDSS3 are smoothed with a 15 pixel boxcar, the spectrum from Magllan/FIRE is smoothed with a 3 pixel boxcar, and the MMT/Binospec spectrum is smoothed with 5 pixel boxcar. The spectrum of J025216.64--050331.8 is from Gemini GMOS-S. All spectra are corrected for Galactic extinction using the \cite{cardelli89} Milky Way reddening law and E(B $-$ V) derived from the \cite{schlegel98} dust map.).
} 
\end{figure}

\section{Spectroscopic Observations} \label{sec:obs}
We conducted spectroscopic observations of these DES-selected candidates with Magellan/FIRE \citep{simcoe08}, Magellan/LDSS3-C \citep{stevenson16} and MMT/Binospec \citep{fabricant98} in May, July and November, 2018. 
FIRE is an IR echelle/longslit spectrograph on the Magellan Baade 6.5 m telescope in Chile.
We used the $1\farcs0$ longslit mode with FIRE, which provides a resolution of $R \sim 300-500$ covering the wavelength from 8000 \AA\ to 22000 \AA\. The throughput at wavelength shorter than 8500 \AA\ is low. We typically use 5 -- 10 minute exposure time for each target. 
LDSS3 is a high efficiency spectrograph and optical imager on the  Magellan Clay 6.5m telescope. We used the VPH-Red grism covering the wavelength range of 6000 - 10000 \AA\ . We used $1\farcs25$ slit with $R \sim 1000$. The typical exposure time for identification is 20 - 30 minutes. 
MMT Binospec is a wide field optical spectrograph capable of mutliobject spectroscopy, single-slit, and imaging with dual 8$' \times$ 15$'$ fields of view. We used the 270 lines/mm grating centered at 7800 \AA\ with a 1.25$''$ slit, which provide a resolution of $R \sim$ 1200 and an effective wavelength coverage from 5130 \AA\ to 10200 \AA\ .
We observed 31 candidates from main sample in total. 
After discovery, we took the follow-up optical spectroscopy for new quasars with LDSS3 and GMOS-S in the Gemini South 8.1m telescope to obtain higher quality optical spectra. 
Data from FIRE, LDSS3 and Binospec was reduced with standard IRAF routines. The GMOS data was reduced using the Gemini IRAF packages.

\section{Results} \label{sec:result}

\begin{deluxetable*}{ l c l l l l l c l l l l}
\tablecaption{Photometric data of six new quasars.}
\tabletypesize{\scriptsize}
\tablewidth{0pt}
\tablehead{
\colhead{Name} &
\colhead{Redshift} &
\colhead{$M_{1450}$} &
\colhead{$z$} &
\colhead{$Y$} &
\colhead{$J$\tablenotemark{a}} &
\colhead{$K$} &
\colhead{NIR} &
\colhead{W1\tablenotemark{b}} &
\colhead{W2}
}
\startdata
DES J021638.85--522620.6 & 6.41 & -25.11$\pm$0.13 & 22.11$\pm$0.06 & 21.72$\pm$0.13 & 20.47$\pm$0.19 &18.95$\pm$0.24& VHS &17.46$\pm$0.05&17.13$\pm$0.15\\
DES J025216.64--050331.8\tablenotemark{c}  & 7.02 & -25.77$\pm$0.09 & 22.52$\pm$0.10 & 20.85$\pm$0.07 & --- & --- &--- &17.52$\pm$0.08&16.98$\pm$0.22\\
DES J024655.90--521949.9 & 6.87 & -25.46$\pm$0.15 & 22.30$\pm$0.15 & 21.24$\pm$0.14 & 20.35$\pm$0.20 & 18.26$\pm$0.18 & VHS & 17.27$\pm$0.04&16.76$\pm$0.11\\
DES J031941.66--100846.0\tablenotemark{c}  & 6.83 & -25.71$\pm$0.20 & 22.35$\pm$0.13 & 21.04$\pm$0.20 & --- & --- & --- & 17.11$\pm$0.04 & 16.12$\pm$0.08\\
DES J213110.29--435902.5 & 6.45 & -25.30$\pm$0.11 & 21.89$\pm$0.05 & 21.54$\pm$0.10  & 20.74$\pm$0.25 & 18.91$\pm$0.24 & VHS & 18.17$\pm$0.13&17.49$\pm$0.30\\
DES J221100.60--632055.8 & 6.88 & -24.93$\pm$0.14 & 22.72$\pm$0.10 & 21.76$\pm$0.13 & 20.60$\pm$0.22 & 19.22$\pm$0.35 & VHS &18.28$\pm$0.14&18.02$\pm$0.44\\
 \enddata
 \tablenotetext{a}{Forced photometric data of $J$ and $Ks$ band image from public survey.}
 \tablenotetext{b}{Forced photometry on unWISE coadds.}
 \tablenotetext{c}{Quasar J025216--050331 and J031941.66--100846.0 do not have public NIR imaging data.}
\end{deluxetable*}

\begin{deluxetable*}{ l r r l l l l l}
\tablecaption{Spectroscopy information of the six new quasars.}
\tabletypesize{\scriptsize}
\tablewidth{0pt}
\tablehead{
\colhead{Name} &
\colhead{Redshift} &
\colhead{$M_{1450}$} &
\colhead{Instrument} &
\colhead{Exptime} &
\colhead{Grating} &
\colhead{Slit} &
\colhead{Obsdate}
}
\startdata
  J021638.85--522620.6 & 6.41 & -25.11$\pm$0.13 & Magellan/LDSS3 & 1800 & VPH-Red & 1.25\_Center & 2018-07-21\\
  J025216.64--050331.8 & 7.02 & -25.77$\pm$0.09 & Magellan/LDSS3 & 1200 & VPH-Red & 1.25\_Center & 2018-07-22 \\
                                             &         &           & Gemini/GMOS-S    &  8000 & R400 & 0.75''longslit & 2018-10-01\\
  J024655.90--521949.9 & 6.87 & -25.46$\pm$0.15 & Magellan/LDSS3 & 1500 & VPH-Red & 1.25\_Center & 2018-07-22\\
  J031941.66--100846.0 & 6.83 & -25.71$\pm$0.20 & MMT/Binospec & 1800 & 270 lines/mm & 1.25'' longslit & 2018-11-08\\
  J213110.29--435902.5 & 6.45 & -25.30$\pm$0.11 & Magellan/FIRE & 600 & --- & 1.0''\_longslit & 2018-05-02\\
  J221100.60--632055.8 & 6.88 & -24.93$\pm$0.14 & Magellan/LDSS3 & 5400 & VPH-Red & 1.25\_Center & 2018-07-21\\
 \enddata
\end{deluxetable*}

At the time of this publication, we have discovered six new quasars at $6.41 < z < 7.02$ from the DES--near-IR--unWISE selection. Four quasars are at $z > 6.8$. 
Quasar J024655.90--521949.9 is also independently discovered by Reed et\ al. (2018. in preparation).
The photometric properties and spectroscopic information of these six new quasars are listed in Table 2 and Table 3. Among other observed candidates, some are L/T dwarfs and some are $'$non-quasar$'$ objects that do not have obvious break and broad emission line but the S/N are too low for definitive identifications.

We measure the quasar redshifts by visually matching the observed spectrum to a quasar template using an eye-recognition assistant for quasar spectra software \cite[ASERA;][]{yuan13}. The matching is based on emission lines $\rm Ly \alpha$ and N\,{\sc v}. The typical uncertainty of the redshift measurement is around 0.03. For spectra of J021638.85--522620.6 and J213110.29--435902.5, due to the low S/N, the uncertainty could be $\sim 0.05$. We do not include the systematic offset of Ly$\alpha$ emission line \citep[e.g.,][]{shen07}, which is typically $\sim$ 500 km/s and much smaller than the uncertainty of template matching. The spectra of the new quasars are shown in Figure 2.

As the spectra of these six new quasars do not have sufficient  wavelength coverage and S/N for reliable continuum fit, we can not measure the $M_{1450}$, absolute continuum magnitude at the rest-frame 1450\AA,  directly from spectra. Therefore, we estimate the $M_{1450}$ using the composite spectra of luminous low redshift quasars \citep{selsing16}. We scale the composite spectrum with Galactic extinction corrected DES $Y$ photometry of each quasar and then measure the magnitude at 1450 \AA\ from the scaled composite spectrum. The $M_{1450}$ magnitudes of the new quasars are also listed in Table 2 and Table 3.

Figure 3 shows the locations of the six new quasars in the $M_{1450}$ -- redshift plot of all known $z > 6.3$ quasars. Compared to Paper III,  the sample presented in this paper extends to fainter population. Although this survey is still on-going and only a small number of candidates have been observed, we have already discovered 5 quasars at $z \gtrsim 6.5$, including four at $z>6.8$, deep into the epoch of reionization. 

{\bf Quasar J025216.64--050331.8} is the fourth known quasar at redshift higher than 7. Its GMOS-S spectrum (Figure 2) shows strong absorption profile close to the Ly$\alpha$ emission line. The Near-IR spectrum is required for the further investigation of this absorption feature, e.g. whether it is due to the Ly$\alpha$ damping wing from a significantly neutral IGM, is due to a broad absorption line feature or the
presence of a proximate damped Ly$\alpha$ absorber). We are collecting the high quality optical and near-IR spectra of the four new $z > 6.8$ quasars for the further studies of their BH masses and IGM properties.

\begin{figure}
\centering
\epsscale{1.2}
\plotone{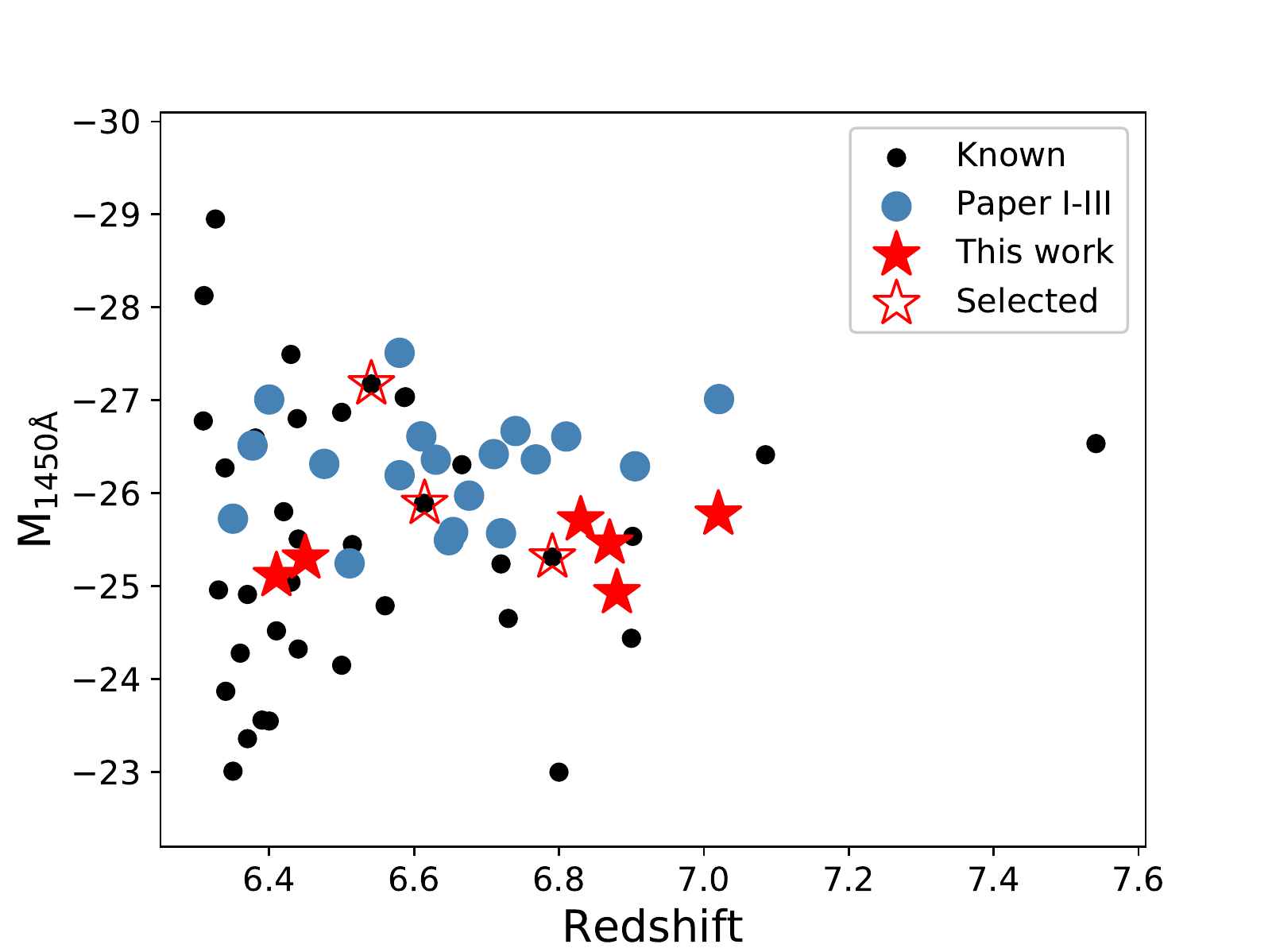} 
\caption{The redshift and $M_{1450}$ distribution of quasars at $z \ge 6.3$. The black filled circles are the previously known quasars. The blue filled circles denote quasars from our surveys in the northern sky \citep[][Paper I, II and III]{wang17,wang18a,wang18b}. The red filled stars represent six new quasars from this survey and the red open stars are the three known quasars in the DES area which meet our selection criteria.
} 
\end{figure}

\section{Discussion and Summary} \label{sec:summary}
The new discoveries, especially the four quasars at $6.83 \le z \le 7.02$, from our DES--near-IR--unWISE quasar selection suggest that the combination of deep DES optical data, VHS/VIKING near-IR and unWISE mid-IR data is highly effective in searching for quasars at $z > 6.5$. Among our six new quasars, only three have ALLWISE detections, three have $J$ band public data and none of them has both $J$ and $W1$ public data used for the $J-$ W1 cut. The forced photometric data of near-IR and unWISE images significantly improve the effectiveness of our photometric dataset for quasar selection. 

With this deep photometric dataset, we are able to reach $\sim$1 magnitude fainter than our survey in the northern sky. Based on the double power quasar luminosity function (QLF) presented in Paper III, we estimate the total number of quasars at $6.5 < z < 7.1$ and $M_{1450} < -24.5$ in the DES area is around 60. 
The quasar sample in Paper III only covers  the relative bright quasars with $M_{1450} < -25.5$ and the LF is measured with fixed faint end slope ($\alpha = -1.90$) and break magnitude ($M^{*} = -25.2$) from the previous $z \sim 6$ QLF \citep{jiang16}. However, the quasar sample from \cite{jiang16} only includes two quasars at $M_{1450} > -24$ and thus the faint end slope and break magnitude are less well constrained. As comparison, we estimate the number based on different QLFs.
Using the single power law fit ($\beta$=--2.35) QLF in Paper III, the estimated number is $\sim$ 67, which can be considered as an upper limit.
\cite{matsuoka18c} report a new measurement of QLF at $z \sim 6$ with a quasar sample covering the magnitude range of $-22 \le M_{1450} \le -29$. For the first time, it shows a clear break in the $z \sim 6$ QLF. This work suggest a flatter faint end slope (--1.23) and a fainter break magnitude (--24.9). We fix $\alpha$ and $M^{*}$ to be --1.23 and --24.9 and re-fit our $z \sim 6.7$ QLF in Paper III. We obtain a bright end slope $\beta = -2.51$ and normalization $\Phi^{*}= 3.69$ Gpc$^{-3}$ mag$^{-1}$. The predicted number of DES quasar based on this new fit is $\sim$ 55.

We have discovered six new quasars among the first thirty candidates that we have observed so far in the DES area. There are a total of 380 candidates. Even though we have not yet modeled the selection completeness and these initial observations tend to focus on relatively bright objects, the rate of our discovery suggests a significant number of additional quasars among the candidates await spectroscopic identifications.  
 
In this paper, we present the initial results of our $z > 6.5$ quasar survey in the DES 5000 deg$^{2}$ area. Compared with our survey in the Paper III, we are constructing a deeper optical/IR dataset by combining the DES optical photometry and VHS, VIKING, ULAS near-IR data and unWISE mid-IR data. We carried out forced photometry of VHS, VIKING, ULAS and unWISE images to improve the survey depth. Although this survey is just starting, the four new quasars at $z > 6.8$ together with quasars from Paper III have already significantly increased the number of quasars at $z > 6.8$. 
With new IR data and future DES data release, we will be able to further expand the survey area and depth. For example, the up-coming deeper version of the unWISE coadds will include the fourth year NEOWISE images. The completed survey will allow measurements of the reionization-era quasar luminosity function measurement to reach one magnitude fainter. 
The quasars discovered in our survey are ideal targets for high quality multiwavelength followup observations to study
the properties of SMBH, IGM and SMBH-host co-evolution.

\acknowledgments
We thank Aaron Meisner and Dustin Lang for the discussions about the unWISE data, and Sophie Reed for discussions regarding DES quasar survey. 
J. Yang, X. Fan and M. Yue acknowledge the supports from the US NSF grant AST 15-15115 and NASA
ADAP Grant NNX17AF28G.
X.-B. Wu and L. Jiang thank the supports by the National Key R\&D Program of China (2016YFA0400703) and the National Science Foundation of China (11533001, 11721303). 

We acknowledge the use of the Magellan Clay and Baade 6.5m telescopes.
This paper also uses data based on observations obtained at the Gemini Observatory, which is operated by the Association of Universities for Research in Astronomy, Inc., under a cooperative agreement with the NSF on behalf of the Gemini partnership: the National Science Foundation (United States), the National Research Council (Canada), CONICYT (Chile), Ministerio de Ciencia, Tecnolog\'{i}a e Innovaci\'{o}n Productiva (Argentina), and Minist\'{e}rio da Ci\^{e}ncia, Tecnologia e Inova\c{c}\~{a}o (Brazil).
This publication makes use of data products from the Wide-field Infrared Survey Explorer, which is a joint project of the University of California, Los Angeles, and the Jet Propulsion Laboratory/California Institute of Technology, and NEOWISE, which is a project of the Jet Propulsion Laboratory/California Institute of Technology. {\it WISE} and NEOWISE are funded by the National Aeronautics and Space Administration. We acknowledge the use of the DES, the VHS, VIKING and ULAS data.

%

\vspace{5mm}
\facilities{DES}, {WISE}, {VISTA}, {UKIRT (WFCam)}, {Magellan Clay (LDSS3)}, {Magellan Baade (FIRE)},{MMT (Binospec)}, {Gemini-South (GMOS)}

\end{document}